\documentclass[12pt]{iopart}
\usepackage{graphicx}
\usepackage{bm}

\def\im{ \,\mathrm{Im}\,}
\def\re{ \,\mathrm{Re}\,}

\begin{document}

\title{ On the correct formula for the lifetime broadened 
superconducting density of states} 

\author{ Bo\v zidar Mitrovi\'c 
and Lee A.~Rozema}

\address{Department of Physics, Brock University,
St.Catharines, Ontario, Canada L2S 3A1}
\ead{mitrovic@brocku.ca}

\begin{abstract}
We argue that the well known Dynes formula  
[Dynes R C {\it et al.} 1978 {\it Phys.~Rev.~Lett.} {\bf 41}  1509]
for the superconducting quasiparticle density of states,  
which tries to incorporate the lifetime broadening in an approximate way,  
cannot be justified microscopically for conventional superconductors.  
Instead, we propose a new simple formula in which  
the energy gap has a finite imaginary part $-\Delta_2$   
and the quasiparticle energy is real.
We prove that in the quasiparticle approximation 
2$\Delta_2$ gives the quasiparticle decay rate at the gap edge for 
conventional superconductors. This conclusion does not depend on the nature of 
interactions that cause the quasiparticle decay. The new formula is tested 
on the case of a strong coupling superconductor Pb$_{0.9}$Bi$_{0.1}$ and 
an excellent agreement with theoretical predictions is obtained.  
While both the Dynes formula and the one proposed in this work give 
good fits and fit parameters for Pb$_{0.9}$Bi$_{0.1}$, only the latter formula 
can be justified microscopically.
\end{abstract}

\submitto{\JPCM}

\maketitle


Almost thirty years ago Dynes, Narayanamurti and Garno \cite{Dynes} proposed  
that the quasiparticle recombination time in a strong-coupled superconductor 
can be directly measured from the width of the peak in the tunneling 
conductance $dI(V)/dV$ of a superconductor-insulator-superconductor tunnel 
junction at the sum of the gaps. They found that the data on 
Pb$_{0.9}$Bi$_{0.1}$-insulator-Pb$_{0.9}$Bi$_{0.1}$ planar 
tunnel junction could be fitted quite well for voltages near 
twice the gap if the quasiparticle density of states  
\begin{equation}
\label{QPDOS}
\rho(E)=\re\frac{E}{\sqrt{E^2-\Delta^2(E)}}\>,
\end{equation}
in the expression for the tunneling current 
\begin{equation}
\label{current}
I(V)\propto\int_{-\infty}^{+\infty}dE\rho(E)\rho(E+eV)
[f(E)-f(E+eV)]
\end{equation}
is replaced by
\begin{equation}
\label{Dynes_formula}
\rho_{D}(E,\Gamma_D)=\re\frac{E-i\Gamma_D}{\sqrt{(E-i\Gamma_D)^2-
\Delta_{0}^{2}}}\>,
\end{equation}
with real and E-independent $\Gamma_D$ and the measured gap edge $\Delta_0$. 
In (\ref{QPDOS}) 
$\Delta(E)$ is the complex gap function and $f$ and $e$ in (\ref{current})
are the Fermi function at temperature $T$ and the   
magnitude of electron charge, respectively. It was proposed 
\cite{Dynes} that the temperature dependent parameter $\Gamma_D$ in 
(\ref{Dynes_formula}) incorporates the quasiparticle lifetime effects. 
A good agreement between the measured $\Gamma_{D}(T)$ and a microscopic
calculation \cite{Dynes} based on the work 
by Kaplan {\it et al.} \cite{Kaplan} for a number 
of temperatures below the transition temperature $T_{c}$ 
of Pb$_{0.9}$Bi$_{0.1}$ 
was taken as a justification for the replacement of 
$\rho(E)$ with $\rho_{D}(E,\Gamma_D)$ and for the interpretation of 
parameter 2$\Gamma_D$ as the inverse of the quasiparticle recombination 
lifetime. 
Formula (\ref{Dynes_formula}) is now widely known as the Dynes 
formula and it has been applied to a variety 
of low temperature ($T\ll T_{c}$) tunneling experiments ranging from  
tunneling into the bulk \cite{Dynes3D}
and thin film \cite{White} inhomogeneous/granular superconductors to 
the tunneling into a two-band superconductor MgB$_2$ \cite{MgB2} 
and tunneling into a novel superconductor CaC$_{6}$ \cite{Bergeal,Kurter}.
The Dynes formula was also recently used to describe the density of states obtained  
in photoemission studies of superconducting h-ZrRuP \cite{Matsui} and 
of filled skutterudite superconductor LaRu$_4$P$_{12}$ \cite{Tsuda}.

However, the {\em ansatz} (\ref{Dynes_formula}) 
cannot be justified  
for a conventional strong coupling superconductor, such as Pb$_{0.9}$Bi$_{0.1}$ 
\cite{Dynes}, from first principles. 
Indeed,  
$\rho(E)$ is given in terms of the diagonal electron Green's function 
in the superconducting state 
\begin{equation}
\label{Green}
G_{11}({\bf k},E)=\frac{EZ({\bf k},E)+\varepsilon_{{\bf k}}}
{E^2 Z^2({\bf k},E)-\phi^2({\bf k},E)-\varepsilon_{{\bf k}}^2}\>,  
\end{equation}
where
$Z$ is the complex renormalization function and $\phi$ is the complex 
pairing self-energy \cite{Schr,Scalapino69}, as 
\begin{equation}
\label{definition}
\rho(E)=-\frac{1}{\pi N(0)}\im\sum_{{\bf k}}G_{11}({\bf k},E)\>,
\end{equation}
where $N(0)$ is the normal state density of states at the Fermi level.
All interactions enter via the self-energy terms $Z$ and $\phi$ 
and assuming that they do not depend on momentum ${\bf k}$ one finds
\begin{eqnarray}
\label{sequence1}
\rho(E)&=&\re \frac{EZ(E)}{\sqrt{E^2 Z^2(E)-\phi^2(E)}}  \\ 
\label{sequence2}
\phantom{\rho(E)}&=&\re \frac{E}{\sqrt{E^2-\Delta^2(E)}}\>,
\end{eqnarray}
where in the last step $Z(E)$ and $\phi(E)$ have been eliminated in 
favor of the gap function $\Delta(E)=\phi(E)/Z(E)$. Clearly, all the 
lifetime effects which enter via $\phi(E)$ and $Z(E)$ are ultimately 
incorporated in the complex gap function $\Delta(E)$ and the tunneling 
current $I(V)$ depends on the full complex gap function as is clear 
from equations (\ref{QPDOS}) and (\ref{current}). Note that (\ref{sequence1})
cannot be cast into the form (\ref{Dynes_formula}) by a suitable choice of 
$Z(E)$ (e.g.~taking $Z(E)=1-i\Gamma_D/E$ would give $\re[(E-i\Gamma_D)/
\sqrt{(E-i\Gamma_D)^2-\phi(E)}]$, where the pairing self-energy $\phi$ 
appears instead of the gap $\Delta$, and the measured $dI/dV$ gives   
$\Delta$ and not  $\phi$).

Instead of replacing $\rho(E)$ with $\rho_{D}(E,\Gamma_D)$ it 
is more reasonable to keep $\Delta(E)$ in (\ref{QPDOS}) constant but 
complex for $E$ not too far from the gap edge $\Delta_0$, i.e.~replace 
(\ref{QPDOS}) with
\begin{equation}
\label{complex-delta}
\rho_{\Delta}(E,\Delta_2)=\re\frac{E}{\sqrt{E^2-(\Delta_0-i\Delta_2)^2}}\>,
\end{equation}
where $-\Delta_2$ is the imaginary part of the gap at $E=\Delta_0$. 
It is well known that at a finite temperature the imaginary part of   
the gap at the gap edge is finite as 
a result of quasiparticle damping (see figure 45 in \cite{Scalapino69}). 
In fact, it is easy to
prove that in the quasiparticle 
approximation \cite{Kaplan} the quasiparticle decay
rate at the gap edge is equal to -2$\im \Delta(E=\Delta_0)$. Assuming
that at $E=\Delta_{0}$ the imaginary parts $Z_2$ and
and $\phi_2$ of $Z$ and $\phi$, respectively, are much smaller than
the corresponding real parts one finds that
\begin{equation}
\label{proof}
-\im \Delta(E=\Delta_0)\approx 
\frac{\Delta_{0}Z_{2}(E=\Delta_{0})-\phi_2(E=\Delta_{0})}{Z_1(0)}\>,
\end{equation}
where $Z_1(0)$ is the real part of $Z(E=0)$. Expression (\ref{proof}) is
identical to the equation of Kaplan {\em et al.} 
for the quasiparticle decay rate parameter
$\Gamma(E=\Delta_{0})$ \cite{Kaplan} (see equation (5) in \cite{Kaplan}).
This
result is quite general and does not depend on the specific 
interactions leading to quasiparticle damping, i.e. whether it is
the electron-phonon interaction which was considered in \cite{Dynes,Kaplan},
or the dynamically screened Coulomb interaction in the presence of
disorder which was assumed to be the cause of lifetime 
broadening in {\em low} temperature tunneling experiments into 
three-dimensional granular aluminum \cite{Dynes3D}
and quench-condensed two-dimensional films of Pb and Sn \cite{White}. 
All that is required for 
\begin{equation}
\label{theorem}
2\Gamma({\bf k},E=\Delta_{0})=-2\im \Delta({\bf k},E=\Delta_{0})\>,
\end{equation}
to be valid, where 2$\Gamma({\bf k},E=\Delta_{0})$ is the 
inverse quasiparticle lifetime with
${\bf k}$ on the Fermi surface,
is that the imaginary parts of $\phi({\bf k},E)$ and $Z({\bf k},E)$ are much
smaller than their respective real parts near the gap-edge. Needless to say, 
(\ref{theorem}) does not apply to unconventional superconductors characterized 
by $\sum_{{\bf k}\in FS}\Delta({\bf k})=$0, where FS is the Fermi surface, for ${\bf k}$ near 
the gap nodes \cite{Dahm}. 

\begin{figure}
\begin{center}
\includegraphics[angle=0,width=8cm]{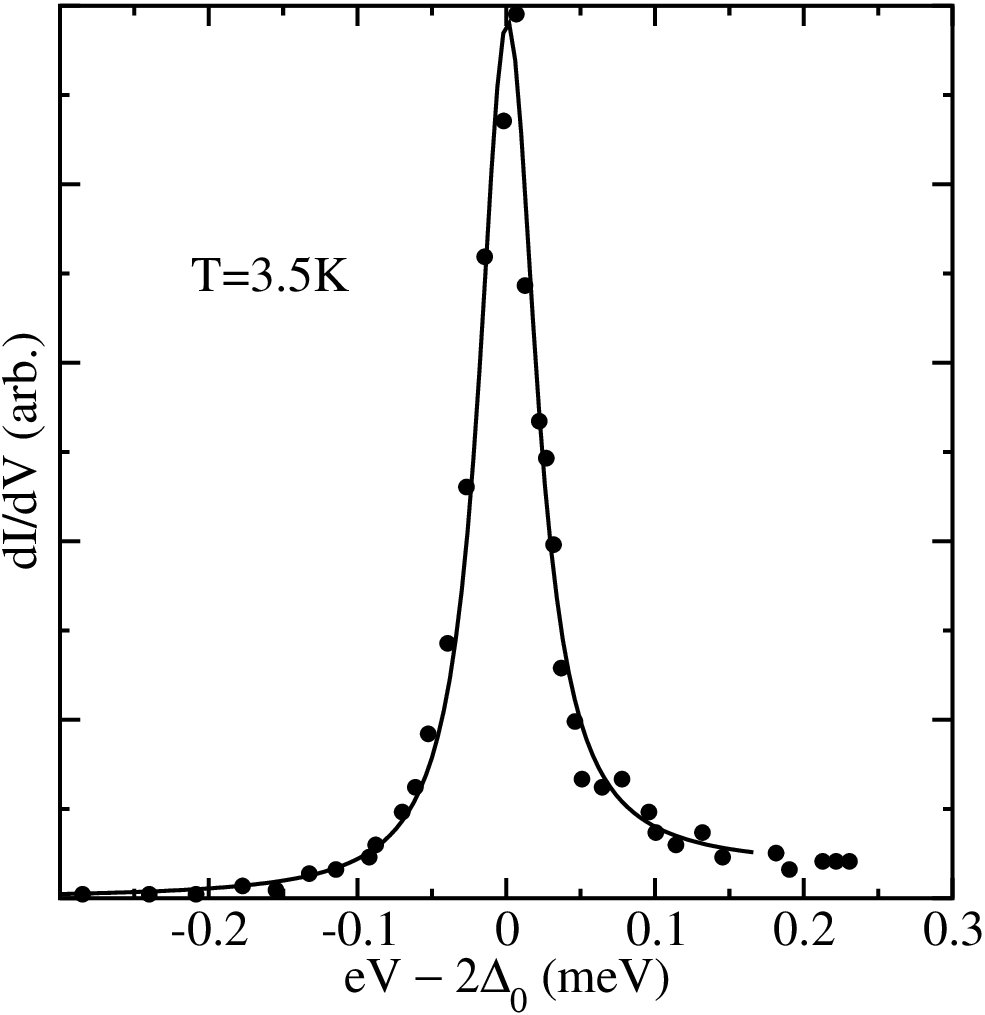}
\end{center}
\caption{The calculated $dI/dV$ (solid line) and the
experimental data points (filled circles) from \cite{Dynes} at $T$=3.5 K.
The data are plotted as a function of $eV$-2$\Delta_{0}$.}
\label{fig:fig1}
\end{figure}

In the case of Pb$_{0.9}$Bi$_{0.1}$ we find that  
equation (\ref{complex-delta}) produces fits to $dI(V)/dV$ which are at least   
as good as 
those obtained with the Dynes formula (\ref{Dynes_formula}). 
Instead of trying to fit the original data from \cite{Dynes}, which 
in addition to the temperature dependent lifetime broadening were
assumed to contain an intrinsic (background) width of 0.01meV, we fitted 
$dI(V)/dV$ calculated from the solutions $\Delta(E)$ and $Z(E)$ of 
the finite temperature Eliashberg equations \cite{Schr,Scalapino69} on the real 
axis using the Eliashberg function 
$\alpha^2(\Omega)F(\Omega)$ for Pb$_{0.9}$Bi$_{0.1}$ \cite{Dynes-Rowell}. 
Thus, the width of the peak in our calculated $dI(V)/dV$ arises 
solely from the temperature dependent lifetime broadening 
and we could compare directly the value of the fit parameter $\Delta_2$ 
in equation (\ref{complex-delta}) 
with our solution $-\im \Delta(E)$ for $E$ at the gap edge. Moreover, we 
could calculate the decay rate parameter $\Gamma(E)$ directly from 
our solutions of Eliashberg equations \cite{Kaplan} (see equation (4) in 
\cite{Kaplan})
\begin{equation}
\label{exact}
\Gamma(E)=EZ_2(E)/Z_1(E)-\phi_1(E)\phi_2(E)/[Z_1^2(E)E]
\end{equation}
and compare its value at $E=\Delta_{0}$ with $\Delta_2$ obtained from 
the fits with equation (\ref{complex-delta}).
We note,
however, that there is a good agreement between the shapes of the calculated  
$dI(V)/dV$ and the measured ones \cite{Dynes} down to $T=$2.75 K as illustrated in 
figure \ref{fig:fig1} for T=3.5 K. In figure \ref{fig:fig1} the results are 
plotted as functions of $eV-2\Delta_0$ since with our choice of 
the Coulomb pseudopotential $\mu^{*}(\omega_{c})$=0.1034, which was 
fitted to the experimental zero temperature gap edge $\Delta_{0}$=1.54 meV 
\cite{Dynes-Rowell} for the cutoff $\omega_{c}$=100 meV in the Eliashberg 
equations, we obtain somewhat higher values of $\Delta_0$ than those 
found in \cite{Dynes}. As the Coulomb pseudopotential term in the 
Eliashberg equations is purely real it does not affect the imaginary parts of 
the solutions \cite{Schr,Scalapino69}.  
 
\begin{figure}
\begin{center}
\includegraphics[angle=0,width=8cm,height=18.5cm]{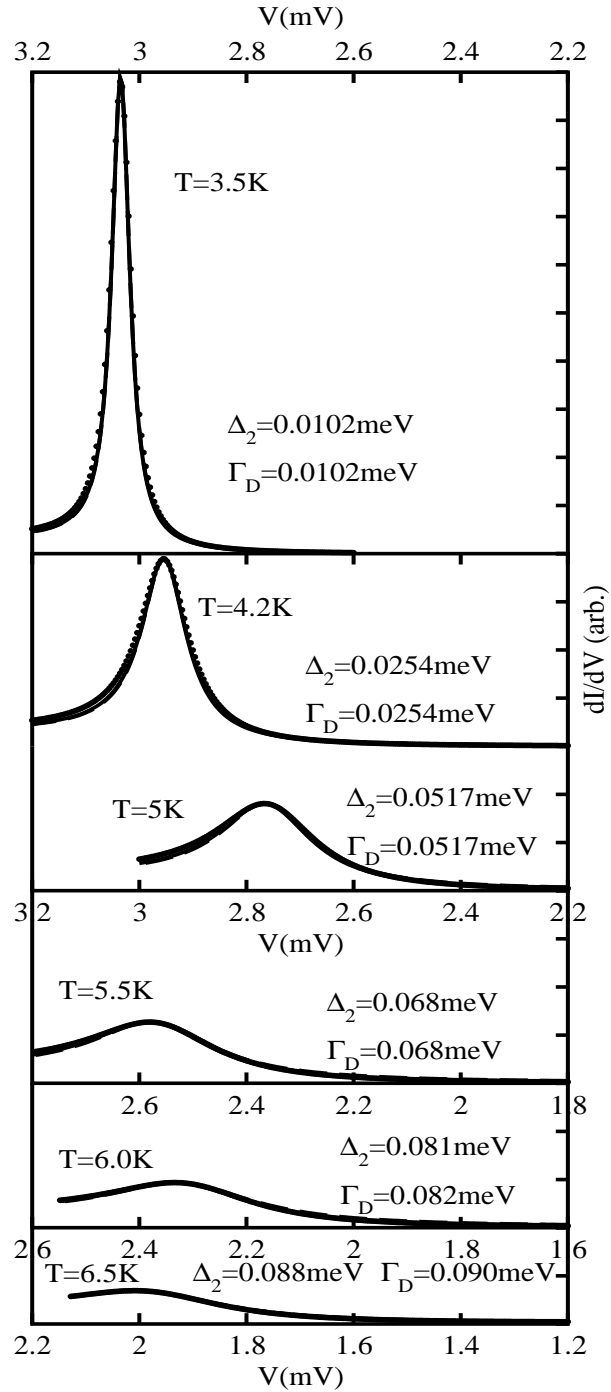}
\end{center}
\caption{The calculated (filled circles) $dI/dV$ at six different
temperatures as a function of voltage and their fits with
(\ref{complex-delta}) (solid line) and with the Dynes formula
(\ref{Dynes_formula}) (dashed line) with $\Delta_2$ and $\Gamma_D$
as the only fit parameters, respectively.}
\label{fig:fig2}
\end{figure}

\begin{figure}
\begin{center}
\includegraphics[angle=0,width=8cm,height=8cm]{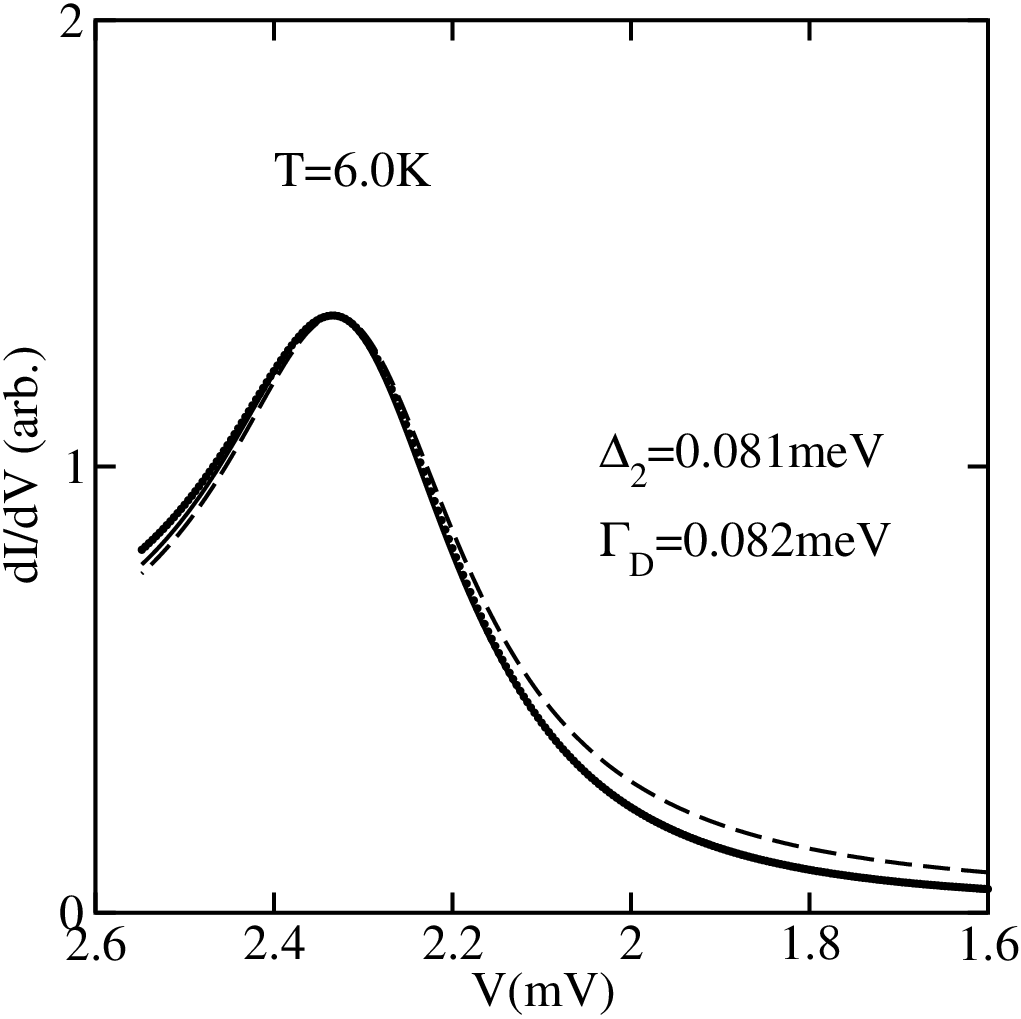}
\end{center}
\caption{The calculated $dI/dV$ (dots) at T=6 K versus voltage
and the fits with (\ref{complex-delta})
(solid line) and (\ref{Dynes_formula})
(dashed line), with $\Delta_2$ and $\Gamma_D$
as the only fit parameters, respectively.}
\label{fig:fig3}
\end{figure}

In figure \ref{fig:fig2} we show the fits to the 
calculated $dI/dV$ using the Dynes formula (\ref{Dynes_formula}) 
and the formula with the complex gap (\ref{complex-delta}). 
On the scale of figure \ref{fig:fig2}, which was chosen to match the 
scale of figure 2 in \cite{Dynes},   
both equations (\ref{Dynes_formula}) and
(\ref{complex-delta}) give equally good fits. Moreover,   
the values of the fit parameter $\Delta_2$ turn out to be nearly the
same as the values of the fit parameter $\Gamma_D$ at all temperatures
considered.  
One can understand why two different functional forms (\ref{Dynes_formula}) 
and (\ref{complex-delta}) give nearly identical fits to $dI/dV$ with 
nearly identical fit parameters $\Gamma_D\approx\Delta_2$ 
from the fact that 
in the limit $\Gamma_D,\Delta_2\ll\Delta_0$ the approximations 
(\ref{Dynes_formula}) and (\ref{complex-delta}) to $\rho(\Delta_0)$ 
give $\sqrt{\Delta_0/\Gamma_D}/$2 and $\sqrt{\Delta_0/\Delta_2}/$2,
respectively and the height of the peak in $dI/dV$ is most sensitive 
to the maximum in the quasiparticle density of states. However, it 
is clear that as the lifetime broadening grows compared to the gap 
edge the difference between the fit parameters obtained with 
(\ref{Dynes_formula}) and with (\ref{complex-delta}) increases and 
the quality of fits with the Dynes formula deteriorates compared to 
the fits with (\ref{complex-delta}) as illustrated in figure \ref{fig:fig3},
in particular at lower voltages. The reason is that for $\Gamma_D,\Delta_2\ll\Delta_0$
in the limit of small energy $\rho_{D}(E,\Gamma_{D})=\Gamma_{D}/\Delta_0$,
while $\rho_{\Delta}(E,\Delta_2)=(\Delta_2/\Delta_{0}^{2})E$
to the first order in $E$, i.e. $\rho_{D}(E,\Gamma_{D})$ does  not vanish at $E$=0.
We note that the experimental {\it low-temperature} densities of states  
obtained for three-dimensional granular aluminum \cite{Dynes3D} do vanish 
at $E$=0 (see figure 3 in \cite{Dynes3D}) , while those obtained for 
two-dimensional quench-condensed tin films \cite{White} do not (see figure 2 in \cite{White}). 
The precise reason for such a difference between three-dimensional and two-dimensional 
disordered conventional superconductors is not known at the present time. 

As one could have expected, the fitted values of $\Delta_2$ turned out to be equal to the 
imaginary parts of our solutions $\Delta(E)$ of the Eliashberg equations 
at $E=\Delta_0$ to within a few percent at all temperatures considered.
The values of $\Delta_2$ extracted from the fits to the calculated $dI/dV$ agree 
with the values of the fit parameter $\Gamma(\equiv\Gamma_D)$ reported 
in \cite{Dynes} before correction for the background to within a percent or two 
down to $T$=4.2 K. At $T$=3.5 K the difference is about 30\% and yet the shapes of 
the calculated and measured $dI/dV$ in figure \ref{fig:fig1} seem to agree quite 
well. A further reduction of the measured $\Gamma$ by the background value of 
0.01meV would increase the difference between the lifetime broadening parameters 
to about 150\%. At $T$=2.75 K our fitted value (the fit is not shown here) is
$\Delta_2$=0.00226 meV which is 80\% 
lower than the measured $\Gamma$ \cite{Dynes} or more than twice the measured 
value after the correction for the background. It is quite plausible that at 
low temperatures, when both the experimental and the theoretical data in the 
peak  
change very rapidly, it is difficult to determine the actual maximum in
$dI/dV$ to which the fit parameters are most sensitive. It is  
likely that the maximum in $dI/dV$ gets underestimated at low $T$ 
having as a consequence
too high values of the lifetime broadening parameter. We believe that is the 
reason for the discrepancies between our  
fitted values of $\Delta_2$ and
those found in \cite{Dynes} at low temperatures and that there is no need to 
invoke the intrinsic temperature-independent broadening parameter.  

\begin{figure}
\begin{center}
\includegraphics[angle=0,width=8cm]{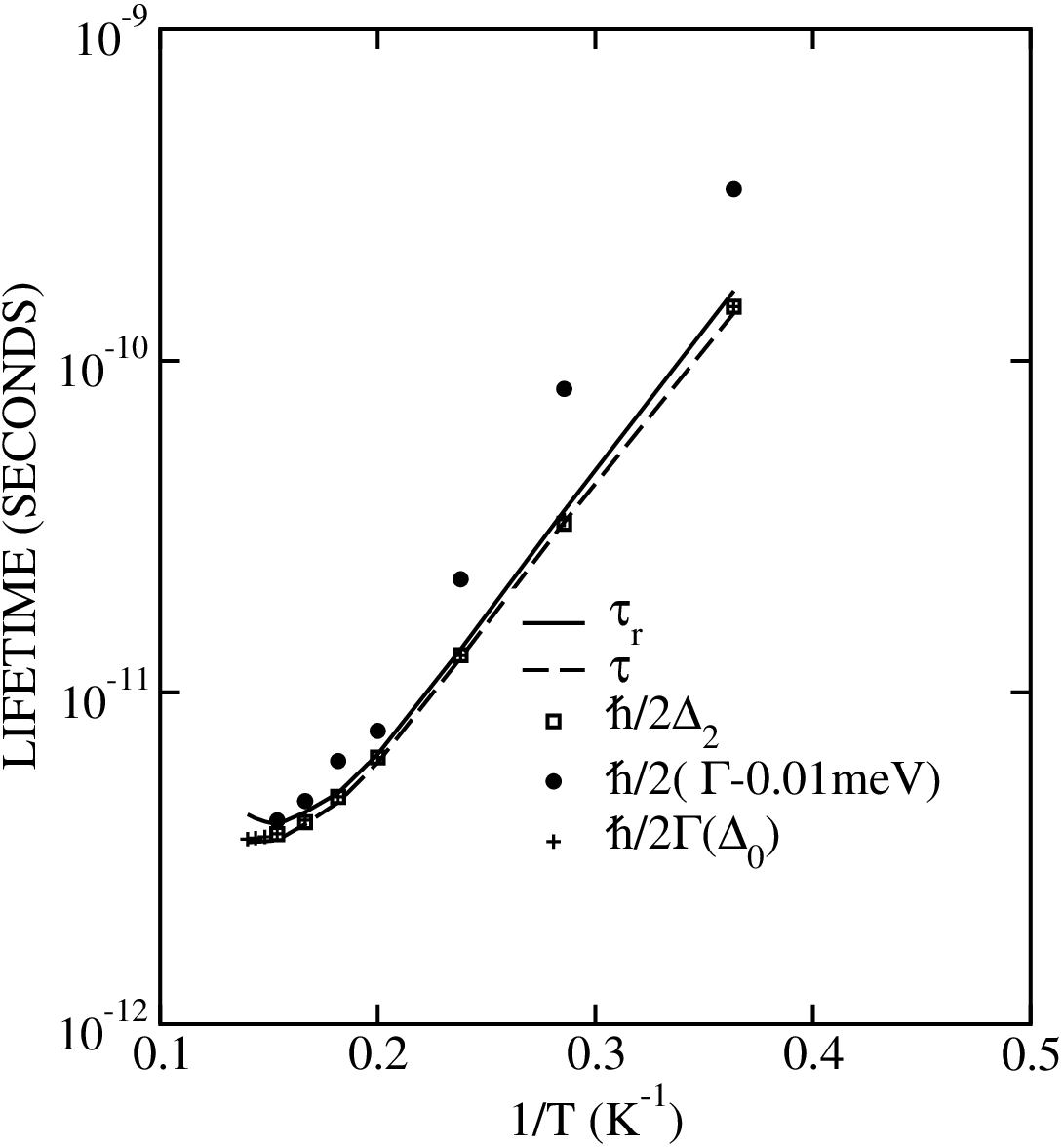}
\end{center}
\caption{ Quasiparticle lifetime at the gap edge
determined in various ways (see the text)
as a function of inverse temperature.
}
\label{fig:fig4}
\end{figure}

Finally, in figure \ref{fig:fig4} we show the temperature dependence of the 
quasiparticle lifetime $\tau$ at the gap edge 
obtained from $\hbar/\tau=2\Delta_2$ 
(open squares) and $\hbar/\tau$=2($\Gamma$-0.01meV) 
(filled circles) with the values of $\Gamma$ taken from  figure 2 in \cite{Dynes}. 
In the same figure we show
theoretical predictions for the recombination time $\tau_{r}$ (solid line) and 
the total lifetime $\tau$ (dashed line) at the gap edge
based on approximate equations
of Kaplan {\em et al.} \cite{Kaplan}
\begin{eqnarray}
\label{recomb}
\frac{\hbar}{\tau_{r}}=C
\int_{2\Delta_{0}}^{\infty}d\Omega\alpha^2(\Omega)F(\Omega) 
\frac{\Omega-\Delta_{0}}{\sqrt{(\Omega-\Delta_{0})^2-\Delta_{0}^2}}
                             \nonumber \\
\times\frac{\Omega}{\Omega-\Delta_{0}}[n(\Omega)+1]f(\Omega-\Delta_{0})\>,\\
\label{total}
\frac{\hbar}{\tau_{s}}=C
\int_{0}^{\infty}d\Omega\alpha^2(\Omega)F(\Omega)
\frac{\Omega+\Delta_{0}}{\sqrt{(\Omega+\Delta_{0})^2-\Delta_{0}^2}}
                             \nonumber \\
\times\frac{\Omega}{\Omega+\Delta_{0}}n(\Omega)[1-f(\Omega+\Delta_{0})]\>,
\end{eqnarray}
where $n(\Omega)$ is the Bose function,
$C=2\pi/\{Z_1(0)[1-f(\Delta_{0})]\}$ and $\hbar/\tau=\hbar/\tau_{r}+\hbar/\tau_{s}$.
A good agreement between the measured $\hbar$/2($\Gamma$-0.01meV) and $\tau_{r}$
calculated according to (\ref{recomb}) was taken as a justification 
of the Dynes formula (\ref{Dynes_formula}) in \cite{Dynes}. 
We note that the integrand in (\ref{recomb}) has a square 
root singularity at the lower limit of integration which has to be handled 
analytically if $\tau_{r}$ is not to be overestimated. Comparing figure 3 in 
\cite{Dynes} and figure 4 in this work it is clear that our calculated $\tau_{r}$
is considerably lower at the low temperatures than the one calculated in \cite{Dynes} 
as the filled circles in both figures represent $\hbar$/2($\Gamma$-0.01meV). 
In addition, we show
in figure \ref{fig:fig4} the lifetime calculated directly from our solutions
of the Eliashberg equations in the quasiparticle approximation
$\hbar/\tau=2\Gamma(\Delta_0)$ (plus signs), where $\Gamma(\Delta_0)$ is
computed using equation (\ref{exact}).
The agreement between the values for the total quasiparticle 
lifetime $\tau$ at the gap edge obtained from the fits with formula
(\ref{complex-delta}) and both theoretical predictions  
is excellent. 

In conclusion, we have shown that one can, indeed, obtain the total quasiparticle 
lifetime at the gap edge from the fits of the derivatives of the $I-V$ 
characteristic of a superconductor-insulator-superconductor tunnel 
junctions using equation (\ref{complex-delta}). The interpretation of the 
parameter 2$\Delta_2$ as the quasiparticle decay rate at the gap edge is 
microscopically justified. While the Dynes formula (\ref{Dynes_formula}) 
gives correct values for the total quasiparticle lifetime, 
it cannot be justified for conventional superconductors.  
Hence the fact that it works, at 
least for the cases when the quasiparticle decay rate is less than 
about 20\% of the gap edge, is a pure accident. It is likely that 
for larger values of $2\Gamma/\Delta_0$, which seems to be the case 
in LaRu$_4$P$_{12}$ ($2\Gamma/\Delta_0\approx$50\%) \cite{Tsuda}, equations (\ref{Dynes_formula}) 
and (\ref{complex-delta}) would give qualitatively and quantitatively 
different results.

\ack

This work
was supported by the Natural Sciences and Engineering Research
Council (NSERC) of Canada. The work of L.~A.~R. was also 
supported in  part through an NSERC Undergraduate Student Research
Award (USRA).

\newpage
~

\newpage

~

\newpage
\newpage


\section*{References}
\begin{thebibliography}{8}

\bibitem{Dynes}
Dynes R C, Narayanamurti V, and Garno J P 1978 {\it Phys.~Rev.~Lett.~} 
\textbf{41} 1509

\bibitem{Kaplan}
Kaplan S B, Chi C C, Langenberg D N, Chang J J, Jafarey S, and
Scalapino D J 1976 {\it Phys.~Rev.~B} \textbf{14} 4854

\bibitem{Dynes3D}
Dynes R C, Garno J P, Hertel G B, Orlando T P 1984 
{\it Phys.~Rev.~Lett.~} \textbf{53}  2437

\bibitem{White}
White A E, Dynes R C, and Garno J P 1986 {\it Phys.~Rev.~B} \textbf{33} 3549

\bibitem{MgB2}
Review issue on MgB$_2$, edited by Crabtree G, Kwok W, Canfield P C 2003
{\it Physica C} \textbf{385} 1

\bibitem{Bergeal}
Bergeal N, Dubost V, Noat Y, Sacks W, Roditchev D, Emery N, H\'erold C,
Mar\^ech\'e J-F, Lagrange P, and Loupias G 2006 {\it Phys.~Rev.~Lett.~} 
\textbf{97} 077003 

\bibitem{Kurter}
Kurter C, Ozyuzer L, Mazur D, Zasadzinski J F, Rosenmann D, Claus H, 
Hinks D G, and Gray K E 2006 {\it Preprint} cond-mat/0612581

\bibitem{Matsui}
Matsui H, Hashimoto D, Souma S, Sato T, Takahashi T, and Shirotani I 2005
{\it J.~Phys.~Soc.~Jpn.~} {\bf 74} 1401

\bibitem{Tsuda}
Tsuda S, Yokoya T, Kiss T, Shimojima T, Shin S, Togasi T, Watanabe S,
Zhang C Q, Chen C T, Sugawara H, Sato H, and Harima H 2006 
{\it J.~Phys.~Soc.~Jpn.~} {\bf 75} 064711

\bibitem{Schr}
Schrieffer J R 1964 {\it Theory of Superconductivity} (New York: W A Benjamin)

\bibitem{Scalapino69}
Scalapino D J 1969 in {\it Superconductivity} ed R D Parks (New York: Marcel Dekker) pp 466-501

\bibitem{Dahm}
Dahm T, Hirschfeld P J, Scalapino D J, and Zhu L 2005 {\it Phys.~Rev.~B} \textbf{72} 214512

\bibitem{Dynes-Rowell} 
Dynes R C and Rowell J M 1975 {\it Phys.~Rev.~B} 1884

\end{thebibliography}
\end{document}